\documentclass[runningheads]{svmult}

\usepackage{makeidx}   
\usepackage{graphicx}  
\usepackage{subeqnar}  
\usepackage{multicol}  
\usepackage{physprbb}  
\makeindex             

\newcommand{\greeksym}[1]{{\usefont{U}{psy}{m}{n}#1}}
\newcommand{\umu}{\mbox{\greeksym{m}}}

\begin{document}
\title*{Infrared Light Curves of Type Ia Supernovae}
%
%
\toctitle{Infrared Light Curves of Type Ia Supernovae}
%
%
\titlerunning{Infrared Light Curves of Type Ia Supernovae}
%
\author{M.~M.~Phillips\inst{1}
\and K.~Krisciunas\inst{2}
\and N.~B.~Suntzeff\inst{2}
\and M.~Roth\inst{1}
\and L.~Germany\inst{3}
\and P.~Candia\inst{2}
\and S.~Gonzalez\inst{1}
\and M.~Hamuy\inst{4}
\and W.~L.~Freedman\inst{4}
\and S.~E.~Persson\inst{4}
\and P.~E. Nugent\inst{5}
\and G.~Aldering\inst{5}
\and A.~Conley\inst{5}}
\authorrunning{M.~M.~Phillips et al.}
%
%
\institute{Las Campanas Observatory, 
           Carnegie Observatories, Casilla 601, 
           La Serena, Chile
\and Cerro Tololo Inter-American Observatory,
     Casilla 603, La Serena, Chile
\and European Southern Observatory, 
     Casilla 19001, Santiago 19, Chile
\and Observatories of the Carnegie Institution of Washington, 
     813 Santa Barbara Street,\\
     Pasadena, CA 91101
\and Lawrence Berkeley National Laboratory, Mail Stop 50-232, 
     1 Cyclotron Road, \\
     Berkeley, CA 94720}

\maketitle              


\section{Introduction}
Beginning in March 1999, we have carried out four major observing campaigns 
with the Swope 1.0~m and du~Pont 2.5~m telescopes at the Las Campanas 
Observatory (LCO) to obtain near-IR light curves of supernovae (SNe) 
of all types.  Data have been obtained for a total of 25 events, with
approximately half of these corresponding to Type~Ia supernovae (SNe~Ia).
These observations are part of a long-term collaborative program at LCO and
Cerro Tololo Inter-American Observatory to investigate in detail the 
photometric properties of SNe~Ia in the near-IR.  These data will be used 
to study the physics of the explosions and to refine the usage of these 
objects as cosmological standard candles.

\section{$JHK$ Light Curve Morphology}
Since the pioneering work of Elias and collaborators\cite{efhp1,emnp1},
the $JHK$ light curves of SNe~Ia have been known to be double-peaked.
The secondary maximum occurs $\sim$30 days after the first maximum,
and is most prominent in the $J$ band.  This double-peaked morphology 
is also observed in the $I$ band\cite{fhrb1}, and is a function of the
decline rate parameter $\Delta$m$_{15}(B)$\cite{phil1} in the sense that
the secondary maximum occurs later and also is generally stronger
in the slowest-declining SNe~Ia\cite{hamuy3,rpk1,kris2}.  Except for the
very fastest-declining events, the primary maximum in $I$ occurs a few days
\emph{before} $B$ maximum\cite{hamuy3}.

\begin{figure}[h]
\begin{center}
\includegraphics[height=0.9\textwidth,angle=-90]{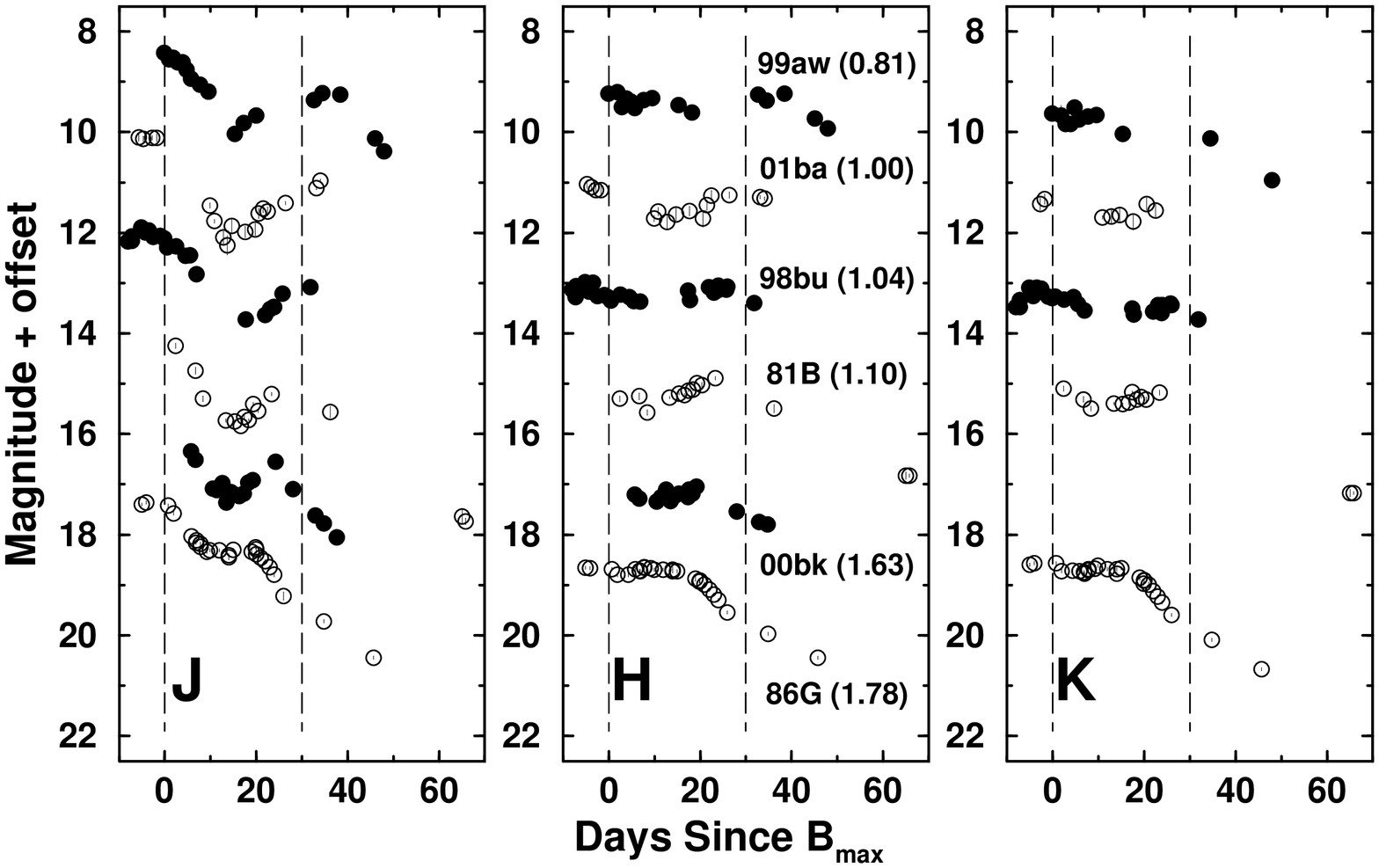}
\end{center}
\caption[]{$JHK$ light curves of SNe~Ia representing a range of decline
rates.  The curves have been arbitrarily shifted in magnitude with respect
to each other for the purposes of comparison.  The value of
$\Delta$m$_{15}(B)$ is given in parentheses after the name of the SN.}
Dashed lines are drawn at 0 and 30 days past $B$ maximum to serve as points 
of reference for the timing of the primary and secondary maxima.
\label{eps1}
\end{figure}

Fig.~\ref{eps1} shows plots of the $JHK$ light curves of six SNe~Ia
covering a range of decline rates.
Data for three of these SNe (1999aw, 2000bk, and 2001ba) were obtained
as part of our observing program; photometry for the other three (1981B,
1986G, and 1998bu) are taken from the 
literature\cite{efhp1,frog1,mpk1,hern1,jha1}.  As in the $I$ band, 
we see that 1) the secondary maximum occurs later for the slower-declining 
events, and 2) the primary maximum occurs a few days before $B$ maximum. 
The dip between the primary and secondary maxima is much less pronounced in
the $H$ and $K$ bands than in $J$, and the secondary maxima reach nearly
the same magnitude as the primary maxima in these two bands.  The net effect
is that the $H$ and $K$ light curves are fairly flat from a few 
days before $B$ maximum to 20-35 days after.

We have also observed a few SNe~Ia in the $Y$ band at 1.035 \umu m\cite{hfpm1}.
At this wavelength -- between the $I$ and $J$ bands -- the secondary maximum 
is extremely prominent and, at least in some cases, brighter than the primary 
maximum.  (The $Z$-band light curve of SN~1999ee shows a similar
behavior\cite{stritz1}.)  We intend to continue collecting data in the 
$Y$~band for future SNe~Ia.

\section{Absolute Magnitudes}
\begin{figure}[h]
\begin{center}
\includegraphics[width=.5\textwidth]{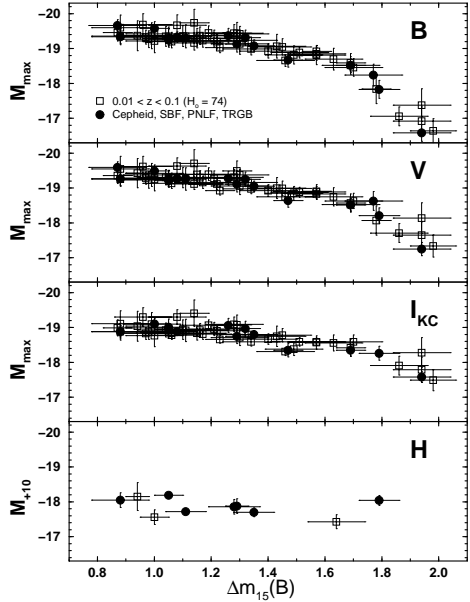}
\end{center}
\caption[]{Absolute magnitudes in $BVIH$ versus $\Delta$m$_{15}(B)$for 
66 well-observed SNe~Ia.}
\label{eps2}
\end{figure}

In the top three panels of Fig.~\ref{eps2} we plot the absolute magnitudes 
in $BVI$ versus the 
decline rate parameter $\Delta$m$_{15}(B)$ for two samples of SNe~Ia.  The
first consists of 16 well-observed nearby SNe~Ia which have occurred in 
host galaxies for which distances have been derived via Cepheids or the
surface brightness fluctuations, planetary nebula luminosity function, or 
``tip of the red giant branch'' methods.  The calibration for all four
methods is based on final results of the HST Key Project to measure the
Hubble constant\cite{freed1}.  The second sample consists of 50 SNe~Ia
in the Hubble flow (i.e. with redshifts greater than 3000 km s$^{-1}$).
Distances for these objects were derived from the host galaxy recession 
velocity assuming H$_0$ = 74 km s$^{-1}$ Mpc$^{-1}$ which gives the best
agreement between the two samples.  The bottom panel shows the
absolute $H$ magnitudes at 10 days after $B$ maximum for the few
SNe in the two samples with $H$-band light curves.  
This epoch was selected rather
than maximum light which occurs several days before $B$ maximum
and therefore has been observed for very few SNe~Ia.  Note that all of the data
plotted in Fig.~\ref{eps2} have been corrected for both Galactic and host
galaxy reddening\cite{phil2}.

In $H$, the slope of the luminosity versus decline rate 
relation appears to be essentially flat for $0.8 < \Delta$m$_{15}(B) < 1.4$.
For the six nearby SNe with independent distance determinations
in this range, we find a weighted mean $H$-band absolute 
magnitude at $t$ = +10 days of M($H_{+10}$) = $-17.91 \pm 0.05$.  
Interestingly, the absolute magnitudes of the two SNe~Ia with 
$\Delta$m$_{15}(B) > 1.4$ are consistent with this value suggesting that
the near-IR luminosities of SNe~Ia may have little or no dependence on 
the decline rate (see 
also \cite{meik1}).  Further observations are obviously required to confirm 
this result.

\section{Hubble Diagram}

\begin{figure}[h]
\begin{center}
\includegraphics[width=.5\textwidth]{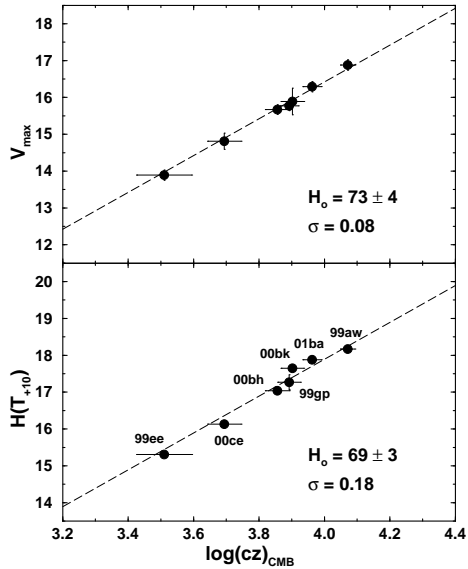}
\end{center}
\caption[]{$V$ and $H$ band Hubble diagrams for seven SNe~I with $z > 0.01$.}
\label{eps3}
\end{figure}

One of the great advantages of working in the near-IR is the minimal effect
of dust reddening (A$_H \sim 0.1$ A$_B$).  If the absolute magnitudes of 
SNe~Ia in the near-IR are essentially independent of the decline rate, these 
objects would then be nearly perfect standard candles at these wavelengths.
Hubble diagrams in $V$ and $H$ for seven SNe~Ia in the Hubble flow 
($z > 0.01$) are shown in Fig.~\ref{eps3}.  For the $V$ band we have corrected
for Galactic and host galaxy extinction, and also for the absolute magnitude
versus decline rate relation\cite{phil2}.  The HST Key Project Cepheid
distances\cite{freed1} are assumed for the calibration of the SNe~Ia absolute 
magnitudes.  A Hubble constant of 73 $\pm$ 4 km s$^{-1}$ Mpc$^{-1}$
is obtained with a small dispersion (0.08 mag), consistent with previous
results\cite{phil2}.

For the $H$-band points in Fig.~\ref{eps3} we again use the magnitudes 
at 10 days after $B$ maximum.
These have been corrected for dust extinction using the same reddenings 
assumed for the $V$ data, but no correction has been made for any possible
dependence of absolute magnitude on decline rate.  A Hubble constant of
69 $\pm$ 3 km s$^{-1}$ Mpc$^{-1}$ is implied assuming the value of M($H_{+10}$)
derived in Sect.~3.
This agrees to within the errors with the value obtained from the $V$ diagram,
although the dispersion of 0.18 mag is a factor of two greater.  There are
at least two (possibly related) reasons why this might be the case: 1) The 
$H$ band measurements were made at 10 days after $B$ maximum
and the scatter at this epoch may be greater due to the 
dependence of the light curve shape on the decline rate, and 2) the assumption
of a constant absolute magnitude as a function of the decline rate may be
incorrect.  These hypotheses can only be tested by obtaining more
near-IR light curves of SNe~Ia which include the maximum-light epoch.
Nevertheless, the consistency between the Hubble constants derived 
independently from the optical and IR data in Fig.~\ref{eps3} 
already suggests that the methods developed for correcting the optical
data for host galaxy reddening and decline rate effects are basically sound.

\section{Interesting Objects}

\subsection{SN~1999ac}

\begin{figure}[h]
\begin{center}
\includegraphics[height=0.55\textwidth,angle=-90]{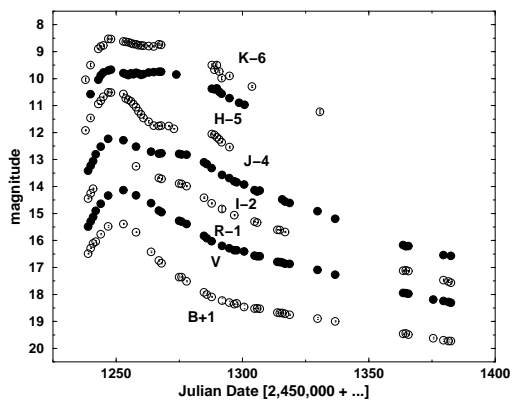}
\end{center}
\caption[]{$BVRIJHK$ light curves of SN~1999ac.}
\label{eps4}
\end{figure}

SN~1999ac was one of the first SN~Ia observed in our program.  As shown in
Fig.~\ref{eps4}, our optical and near-IR light curves begin $\sim$2~weeks
before $B$ maximum, making them among the most detailed ever obtained of a 
SN~Ia.  Spectra taken before maximum indicated that this
SN was a ``peculiar'' event similar to SNe 1991T and 1999aa\cite{phil3}.
The optical light curves of SN~1999ac were also peculiar~--~in particular, the
$B$ light curve displayed a slow rise to maximum similar to the light curve of 
SN~1991T, but then declined much more rapidly than 1991T.  These photometric
peculiarities make it difficult to estimate the host galaxy reddening via
standard techniques.

\subsection{SN~2001ay}

\begin{figure}[h]
\begin{center}
\includegraphics[height=0.6\textwidth,angle=-90]{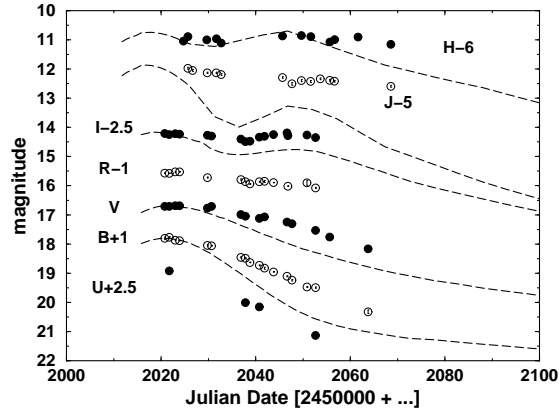}
\end{center}
\caption[]{$UBVRIJH$ light curves of SN~2001ay. Dashed lines show template
light curves of a ``typical'' SN~Ia.}
\label{eps5}
\end{figure}

SN~2001ay appeared in the early-type spiral galaxy IC~4423.  A spectrum
obtained near maximum indicated that the ejecta had unusually 
high expansion velocities\cite{mjck1}.  Fig.~\ref{eps5} shows our $UBVRIJH$
light photometry compared with template curves for a ``typical''
SN~Ia.  The decline rate of 2001ay ($\Delta$m$_{15}(B) \sim$ 0.6-0.7) is the
slowest that we have ever observed.  Note the plateau character of the $I$
and $J$ light curves with no strong dip between the two maxima.  It is not
clear if this photometric peculiarity is related to the high ejecta velocities,
or is simply a characteristic of all SNe~Ia with such slow decline rates.
As in the case of SN~1999ac, we cannot use standard methods to derive the
host galaxy extinction of SN~2001ay, but this is likely to have been fairly
small (E($B-V$)$ \leq 0.1$) since an echelle spectrum\cite{nugent1} showed 
Na~I~D lines at
the redshift of IC 4423 which are no stronger than those due to the Milky Way.
This implies M($V_{max}$) $\sim -19.2$, which
is surprisingly low for such a slow-declining SN~Ia.  On the other hand, in
the near-IR this object was significantly more luminous 
(M($H_{+10}$) $\sim -18.6$) than the other SNe~Ia we have observed to date
(see Sect.~3).

%

\end{document}